# Superconductivity at 22.3 K in Compressed Sodium-intercalated Graphite

Ming-Xing Huang[1,2]†, Yuan-Qing Liu[1,3]†, Chun-Mei Hao[1,2]†, Xi Shao[2], Tingwei An[2], Guochun Yang[2], Yufei Gao[1], Shaojie Wang[1], Lin Wang[1], Bo Xu[1], Feng Ke[1*], Xiang-Feng Zhou[1,2,3*], and Yongjun Tian[1*]

[1]*Center for High Pressure Science, State Key Laboratory of Metastable Materials Science and Technology, Yanshan University, Qinhuangdao 066004, China*
[2]*Hebei Key Laboratory of Microstructural Material Physics, School of Science, Yanshan University, Qinhuangdao 066004, China*
[3]*Key Laboratory of Weak-Light Nonlinear Photonics, School of Physics, Nankai University, Tianjin 300071, China*

†These authors contributed equally to this work

*Corresponding authors: fengke@ysu.edu.cn, xfzhou@ysu.edu.cn, fhcl@ysu.edu.cn

**Abstract**

Graphite intercalation compounds (GICs) have long been recognized as promising candidates for high-temperature superconductivity by intercalation or charge doping, yet experimental progress has stalled with transition temperatures ($T_c$) limited to 11.5 K at ambient pressure and 15.1 K at 7.5 GPa in calcium-intercalated graphite over decades. Here, we report robust superconductivity in sodium-intercalated graphite with $T_c$ of 22.3 K, as demonstrated by clear zero-resistance behavior. Our approach involves simply room-temperature grinding of graphite with sodium, followed by slight compression up to 7.1 GPa, circumventing complex synthesis procedures. Through synchrotron X-ray diffraction combined with first-principles calculations, we identify the major superconducting phase as an orthorhombic stage-2 GIC structure with slightly over-stoichiometric composition ($Na_{1+\delta}C_8$). Electron-phonon coupling calculations reveal that superconductivity primarily emerges from the interactions between out-of-plane carbon π electrons and low-frequency Na/C vibrations. The enhancement in $T_c$ establishes sodium as superior for achieving higher-$T_c$ in GICs and illustrates promising pathway for further optimization through compositional and structural tuning.

**Introduction**

Carbon-based superconductors hold unique promise for technological applications due to their lightweight nature, material abundance, and potential for high-temperature superconductivity (*1-3*). Within this domain, GICs represent a distinctive class of layered structures in which foreign atoms or molecules (guests) are inserted between graphene sheets (hosts) in period arrangements along graphite crystallographic *c*-axis (*4*). This structural architectural creates a superlattice with remarkable configurational versatility, allowing for diverse stacking sequences and stoichiometries that fundamentally change the electronic landscape (*5-8*). The resulting compounds often exhibit superconductivity, characterized by zero electrical resistance and perfect diamagnetism, despite neither the pristine graphite nor the intercalant metals exhibit such behavior independently. This emergent phenomenon has attracted sustained scientific interest, both for the fundamental questions it raises about superconducting mechanisms and for the tantalizing prospect for achieving high-temperature superconductivity in carbon-based systems.

The superconducting narrative of GICs began in the mid-1960s with the discovery of a transition temperature ($T_c$) of approximately 0.55 K in $KC_8$, a potassium-intercalated graphite (*9*). This initial breakthrough catalyzed systematic exploration to elevate $T_c$ through strategic intercalation with various alkali and alkaline earth metals (*9-19*), and through application of external pressure (*16-18*). For decades, however, $T_c$ values remained frustratingly low until significant advantages emerged with ytterbium and calcium intercalation ($YbC_6$ and $CaC_6$), yielding $T_c$ values of 6.5 and 11.5 K, respectively, at ambient pressure (*20, 21*). Subsequent high-pressure investigations raised $CaC_6$'s $T_c$ to 15.1 K at 7.5 GPa (*22, 23*), establishing a record that has stood unchallenged for nearly two decades. Parallel research has extended to graphene—the fundamental building block of graphite—demonstrating superconductivity through either chemical doping/intercalation with alkali and alkaline earth metals (*2, 24, 25*) or through twist-angle engineering (*1, 26-28*), though these approaches have yet to surpass the $T_c$ values achieved in their GIC counterparts.

The superconducting mechanism in $CaC_6$ is well-described by conventional electron-phonon coupling model (*23, 29-32*), with dominant contributions arising from the interactions between electrons at the Ca-derived Fermi surface and phonons associated with Ca's in-plane and C's out-of-plane vibrations. According to the BCS theory (*33*), superconducting $T_c$ in GICs could potentially be enhanced by substituting Ca with lighter elements such as Na and Li, as materials composed of lighter elements typically exhibit high Debye temperatures (*34, 35*), one of critical parameters governing $T_c$. Paradoxically, however, previously reported Li- and Na-based GICs host only modest $T_c$ values not exceeding 5 K even under pressure, *e.g.*, 1.9 K in LiC$_2$ (*18*), 5.0 K in NaC$_2$ (*17, 19*), and 3.8 K in NaC$_3$ (*17*). This striking contradiction between theoretical expectations and experimental observations has presented a formidable challenge for synthesizing GICs with $T_c$ values beyond the 15.1 K record established by compressed CaC$_6$. Encouragingly, recent *ab initio* calculations have predicted that Na-intercalated graphite (NaC$_4$-GIC) could potentially achieve substantially higher $T_c$ (*i.e.*, 41 K at 5 GPa and 48 K at 10 GPa) as determined through solutions of isotropic and anisotropic Migdal-Eliashberg equations (*36, 37*), suggesting a viable pathway toward surpassing current $T_c$ limitation in GICs.

**Results**

**Superconductivity and Hall effect in compressed Na-intercalated graphite (Na-GIC)**

In this work, we report that Na-GIC hosts superconductivity with $T_c$ of 22.3 K at 7.1 GPa, elevating the $T_c$ record for GICs above the boiling temperature of liquid hydrogen. Superconducting Na-GIC was obtained using a two-step synthesis methodology. Firstly, we synthesized Na-GIC precursors by simply grinding the mixture of Na and graphite powders in an inert atmosphere at ambient conditions. The ambient-pressure XRD pattern of the post-grinding product (fig. S1) resembles that of NaC$_{64}$ (*38*). We then compressed the as-synthesized sample to a few GPa to enter the superconducting state in a diamond anvil cell (DAC). This method circumvents traditional challenges associated with alkali metal intercalation, which typically require elevated temperatures or complex multi-step procedures (*17*). Fig. 1A presents the temperature-dependent electrical resistance of Na-GIC at 7.1 GPa. The resistance exhibits metallic behavior (d$R$/d$T$ > 0) upon cooling, followed by a sharp decrease beginning at 22.3 K and ultimately reaching zero resistance at 14.5 K, a hallmark signature of superconducting transition. Multiple experimental runs yield consistent results (fig. S2), with only minor variations in the precise pressure and observed $T_c$(onset) (defined as the onset temperature of the superconducting transition). The response of this phenomenon to external magnetic field provides definitive confirmation of superconductivity. As shown in Fig. 1B, increasing magnetic fields gradually suppress the transition temperature, shifting toward lower values while keeping the normal-state resistance above the transition unchanged. This behavior distinguishes veritable superconductivity from conventional magnetoresistance effect. The achievement of zero resistance combined with systematic magnetic field suppression unambiguously establish the superconducting nature of Na-GIC. From the magnetic field dependence of the transition temperature, we determine the upper critical field $H_c(0)$ and GL coherence length ($\xi_{GL}$) using the Ginzburg-Landau equation (39):

$$H_c(T) = H_c(0) \frac{1-(T/T_c)^2}{1+(T/T_c)^2}, \quad H_c(0) = \phi_0/2\pi\xi_{GL}^2, \tag{1}$$

where $\phi_0 = 2.07\times10^{-15}$ T·m$^2$ is the magnetic flux quantum (fig. S3). This analysis yields $H_c(0)$ = 11.15±0.12 T and a Ginzburg-Landau coherence length $\xi_{GL}(0)$ = 5.44±0.11 nm. Fig. 1C summaries the previously reported $T_c$ in GICs. Notably, the observed $T_c$(onset) of 22.3 K at 7.1 GPa in the as-synthesized Na-GIC exceeds the previous highest value for sodium-based GICs (approximately 5 K in NaC$_2$ at 3.5 GPa) by more than fourfold and pushes the $T_c$ record for GICs above the boiling temperature of liquid hydrogen.

We systematically investigated the pressure dependence of superconductivity in Na-GIC (Fig. 2 and fig. S2), and found that the superconducting transition exhibits remarkable pressure sensitivity while maintaining almost reversible behavior during compression-decompression cycles. Initial signature of superconductivity emerges at 6.0 GPa with a $T_c$(onset) of approximately 8.0 K (Fig. 2A). With further compression up to 7.3 GPa, $T_c$(onset) increases dramatically by a factor of about 2.8, reaching a maximum of 22.3 K. Beyond this pressure, $T_c$ decreases sharply to 8.1 K at 11.0 GPa (the highest pressure in our DAC measurements). During decompression (Fig. 2B and 2C), the superconducting transition is almost reversible with $T_c$ reaching 21.9 K at 6.6 GPa and 10.1 K at 6.0 GPa. Below 5.5 GPa, superconductivity vanishes over the entire temperature range (1.8–300 K). Upon subsequent recompression to 5.8 GPa, superconductivity re-emerges at approximately 9.7 K, albeit with a broader transition width between $T_c$(onset) and $T_c(R\approx0)$ compared to that of the initial compression cycle. This reversible appearance and disappearance of superconductivity with pressure cycling confirms that the observed behavior is an intrinsic property of Na-GIC rather than an artifact of extrinsic factors such as grain cracking, electrode contact issues, or grain-boundary effects, which typically produce irreversible changes. We note minor variations in normal-state resistances ($R_0$) during pressure cycles (*e.g.*, $R_0 \approx 2.7$ mΩ at 6.0 GPa of initial compression versus $R_0 \approx 3.9$ mΩ at 5.8 GPa of recompression), which may be attributed to these extrinsic factors.

Figure 2D presents the pressure-dependent carrier concentration of Na-GIC derived from Hall effect measurements conducted at 30 K. The Hall resistance versus magnetic field curves exhibit nearly linear behavior with negative slope across all measured pressures (fig. S4), indicating electron-dominated transport in the compressed Na-GIC. Superconductivity emerges at electron concentrations of $(1.1–1.6) \times 10^{22}$ cm$^{-3}$ at 6.0 GPa and reaches optimal conditions at $(2.1–2.7) \times 10^{22}$ cm$^{-3}$ at 7.1 GPa. At higher pressures, the electron concentration saturates at approximately $3.0 \times 10^{22}$ cm$^{-3}$, remaining essentially constant up to 11.0 GPa, the highest pressure of this study.

**Structure of superconducting Na-GIC**

To elucidate the crystallographic structure of superconducting Na-GIC, we performed synchrotron X-ray diffraction (XRD) measurements on powdered samples at representative pressures (Fig. 3A). An obvious pressure-induced structural evolution occurs in Na-GIC that correlates with the emergence and optimization of superconductivity. Beginning at 3.4 GPa, a characteristic diffraction peak (marked by blue arrow) that typically represents a new stage of intercalation emerges at approximately 7.50° ($d \approx 3.695$ Å). Its intensity increases substantially with further compression. At 5.3 GPa, an additional characteristic peak of intercalation appears at about 3.77° ($d \approx 7.348$ Å). The XRD patterns are similar from 5.3 to 7.3 GPa, except slight changes in the peak intensities, indicating the completion of this sluggish structural transition by 5.3 GPa. At 8.3 GPa and above, two additional peaks emerge at approximately 6.24° and 8.63°, concurrent with the disappearance of the 3.77° peak and sharp intensity decrease of the 7.50° peak, signifying the formation of other high-pressure phase. The correlation between structural and superconducting transitions is striking: the completion of the former structural transition at 5.3 GPa coincides closely with the onset of superconductivity at 5.8 GPa, while the subsequent structural transition at 8.3 GPa corresponds to the sharp reduction of $T_c$ above 8 GPa. These correlations reveal an intimate relationship between the crystal structure and superconducting properties within this narrow pressure range for our Na-GICs.

The stacking stage of the synthesized Na-GIC can be estimated using the characteristic (00*l*) peak from XRD patterns, as established in previously published studies (*4, 40, 41*),

$$d_{(00l)} = I_c / l, \text{ and } d_{(00l+1)} = I_c / (l + 1) \quad (2)$$

where $d_{(00l)}$ and $d_{(00l+1)}$ are d-spacings of the (00*l*) and (00*l*+1) reflections, and $I_c$ is the *c*-axis repeating distance, respectively. Based on the characteristic peaks of $d \approx 7.348$ Å and 3.695 Å at 7.3 GPa (Fig. 3A), we obtained $l / (l+1) = d_{(00l+1)} / d_{(00l)} = 3.695/7.348 \approx 1/2$. Thus, $l \approx 1$ and $I_c = d_{(001)} \times 1 \approx 7.348$ Å. The $I_c$ value is approaching to that of stage-2 Na-GIC (~7.85 Å, section I, Supplementary Materials). To determine the crystal structure, we then simulated XRD patterns for all the thermodynamically stable stage-2 Na-GICs (*36, 37*), and found that a stage-2 NaC$_8$ structure with *Pmma* symmetry (*37*) provides the best match to our experimental data (fig. S5), particularly for the characteristic diffraction peaks at $d \approx 3.695$ and 7.348 Å that correspond to the altered interlayer spacing of graphite due to Na intercalation. Le Bail refinement using GSAS software (Fig. 3B) yields lattice parameters of $a = 4.507\pm0.011$ Å, $b = 7.367\pm0.016$ Å, and $c = 4.846\pm0.011$ Å at 7.3 GPa, in good agreement with calculated values of NaC$_8$ at 7 GPa ($a = 4.286$ Å, $b = 7.190$ Å, and $c = 4.946$ Å). In this structure, carbon atoms occupy the 8*l* Wyckoff position with coordinates (0.417, 0.699, 0.624) and (0.917, 0.296, 0.875), while sodium atoms occupy the 2*e* position with coordinate (0.250, 0.000, 0.796). Different from the AB-stacking sequence in graphite, the graphene layers in NaC$_8$ are AA-stacking. Under compression, sodium intercalation significantly alters the electronic structure through substantial charge transfer from sodium atoms to graphene sheets. This charge redistribution modifies the electrostatic environment and van der Waals interactions between layers, ultimately favoring AA- over conventional AB-stacking (fig. S6). We note that several weak diffraction peaks at ~10.1°, ~10.5° and ~11.0° deviate from the NaC$_8$ structure (fig. S5). The ~10.5° peak is from the strongest (110) peak of residual Na. Interestingly, a stage-2 NaC$_6$ structure with *P2/m* symmetry (fig. S5 and section II, Supplementary Materials) was found to match both the impurity peaks (~10.1° and ~11.0°) and the characteristic stage-2 Na-GIC peaks (~3.77° and ~7.50°). Incorporating both phases in our model yields substantially better agreement with the experimental data (Fig. 3B). This suggests the coexistence of the major

*Pmma* and minor *P*2/*m* phases in the measured sample.

**Electronic structure and superconducting mechanism of Na-GIC**

To unveil the underlying mechanism of superconductivity in Na-GIC, we performed detailed electron-phonon coupling (EPC) calculations on the identified structure of NaC$_8$ (Fig. 4). Initial calculations on stoichiometric NaC$_8$ yield an EPC strength $\lambda = 0.21$ and a logarithmic average phonon frequency $\omega_{\log} = 1073.96$ K at 7 GPa (figs. S8 and S9). Using the Allen–Dynes modified McMillan equation (*42, 43*) with Coulomb pseudopotential $\mu^* = 0.1$, a negligible $T_c$ of 0.002 K is predicted, which is significantly lower than our experimental observations. Given the presence of excess Na in the as-synthesized sample, we hypothesized a slightly over-stoichiometric composition of Na$_{1+\delta}$C$_8$ (where δ represents additional Na content) for the superconducting phase. Systematical calculations reveal that an optimized electron doping of 0.02 *e*/atom dramatically enhances the superconducting properties while maintaining the structural and dynamically stability (Fig. 4B and Table S1). The doped structure essentially keeps the original lattice parameters and retains metallic with four bands crossing the Fermi surface at 7 GPa (Fig. 4A). Electronic structure analysis reveals that states near the Fermi surface are predominantly derived from carbon's out-of-plane $p_y$ orbitals (perpendicular to the graphene layers, Fig. 3C) with minor contributions from sodium *p* states. For the electron-doped structure, a substantially enhanced EPC constant $\lambda = 0.93$ is obtained, which is more than threefold larger than that of stoichiometric NaC$_8$ at 7 GPa. Accordingly, a $T_c$ of 10.88 K is obtained using the Allen–Dynes modified McMillan equation. Integration of the Eliashberg phonon spectral function $\alpha^2F(\omega)$ indicates that approximately 66.7% of the total EPC strength $\lambda$ arises from low-frequency (below 400 cm$^{-1}$) Na/C vibrations, with the remaining 33.3% attributed to mid- and high-frequency carbon vibrations (Fig. 4B). These results demonstrate that coupling between out-of-plane carbon π-electrons and low-frequency Na/C vibrations dominate the superconducting mechanism in doped NaC$_8$.

The layered structure of Na$_{1+\delta}$C$_8$ resembles other anisotropic superconductors like MgB$_2$, prompting us to investigate potential superconducting anisotropy through self-consistent solutions of the fully anisotropic Migdal-Eliashberg equations (*44, 45*). Band structure calculations using maximally localized Wannier functions (*46*) show excellent agreement with direct density functional theory results (fig. S10), validating our computational reliability. At 5 K, both anisotropic EPC coefficient ($\lambda_{n\mathbf{k}}$) and superconducting energy gaps ($\Delta_{n\mathbf{k}}$), calculated with $\mu^* = 0.1$, exhibit pronounced anisotropy across the Fermi surfaces formed by the four bands crossing the Fermi level (fig. S11). The strongest coupling occurs near the $\Gamma$ point in the Brillouin zone, primarily associated with out-of-plane carbon $\pi$ electrons. The renormalized superconducting density of states at representative temperature (5 K) display a single peak, indicating single-band superconducting behavior in $Na_{1+\delta}C_8$ (Fig. 4C). $\Delta_{n\mathbf{k}}$ varies substantially from 0.84 to 4.08 meV with temperature (Fig. 4D). By analyzing the temperature dependence of $\Delta_{n\mathbf{k}}$, a $T_c$ of 20.8 K is obtained, which is in agreement with our experimental observation of 22.3 K. This close correspondence between theory and experiments provides compelling evidence that conventional EPC is indeed the underlying mechanism for the enhanced superconductivity in this system. Our findings change the landscape of carbon-based superconductors in several significant ways. First, we demonstrate that sodium, historically overshadowed by calcium in the pursuit of high-$T_c$ GICs, actually offers superior performance when properly structured and pressurized. Second, our straightforward room-temperature synthesis approach circumvents traditional challenges associated with alkali metal intercalation (*17*), potentially enabling broader exploration of related systems.

**Discussion**

In conclusion, we have demonstrated superconductivity with a critical temperature of 22.3 K in Na-GIC at a few GPa pressure, pushing the $T_c$ record for GICs above the boiling temperature of liquid hydrogen. The identified superconducting $Na_{1+\delta}C_8$ adopts an orthorhombic stage-2 structure where superconductivity originates primarily from coupling between out-of-plane carbon $\pi$ electrons and low-frequency Na/C vibration modes. This mechanistic understanding opens promising avenues for further optimization through precise control of electron doping, temperature annealing treatment, intercalant distribution, and interlayer spacing, potentially yielding even higher transition temperatures or ambient-pressure analogs, reinvigorating the pursuit of high-temperature superconductivity in lightweight carbon-based materials.

## Materials and Methods
### Synthesis of Na-GIC sample

Graphite powders (Alfa Aesar, 99.99%, 200 mesh) and sodium metal (Alfa Aesar, 99.9%) were used as the raw materials for synthesis of Na-GIC. The Na and graphite raw materials (mixed with molar ration of 1:6) were grinding using an agate mortar for 30 minutes at room temperature. All the entire processes were conducted in a glovebox filled with argon gas ($O_2$ < 0.01 ppm, $H_2O$ < 0.01 ppm). After grinding, the color of sample changes from the shiny Na metal and black graphite to gray, suggesting the formation of Na-GIC, consistent with the ambient-pressure XRD results (fig. S1). It is noticed that a certain amount of Na metal and graphite remain unreacted, which mixed with the gray samples. We ignored these shiny parts of sample and selected the gray Na-GIC as the starting materials for further measurements. All the samples were saved, and loaded into a diamond anvil cell in an argon-filled glovebox before measurements.

### Resistance and Hall measurements

A home-made nonmagnetic Be-Cu cell equipped with two pieces of IIas diamonds with culet diameter of 300 μm were used for high-pressure resistance and Hall effect measurements. An insulating gasket was prepared from a pre-indented T301 stainless steel sheet (~250 μm in thickness) entirely enveloped by a compact insulating layer made of a mixture of epoxy and cubic boron nitride (1:10 weight ratio). A hole with diameter of 120 μm was drilled and used as the sample chamber. The detailed preparation process of an insulating gasket can be found in our previous studies (*47-49*). Gray-colored Na-GIC powders, along with a ruby ball as pressure calibrant, were loaded into the sample chamber in an Ar-filled glovebox. No pressure-transmitting media were loaded to ensure good contact with probing electrodes and preserve the sample from possible reaction with the loaded pressure media. Four platinum (Pt) foil electrodes in a van der Pauw configuration were prepared by a hand-wiring method. The assembled diamond anvil cell was subsequently put into a Physical Property Measurement System (DynaCool, Quantum Design) for measurements at low temperatures (1.8 – 300 K) and under high magnetic fields (-5 – 5 T). For Hall effect measurements, the magnetic fields (-5 –

5 T) were applied perpendicular to the surface of diamond anvil cell.

**Structure characterization at ambient and high pressure**

Powder X-ray diffraction (XRD) at ambient conditions was performed using a Cu $K_\alpha$ radiation with $\lambda$ = 1.541 Å (Bruker D8 Advance diffractometer). Powdered Na-GIC samples were sealed in capton tapes to protect the sample from oxidation. High-pressure synchrotron XRD measurements were carried out at the BL17U1 beamline of Shanghai Synchrotron Radiation Facility (SSRF) with $\lambda$ = 0.4834 Å. The X-ray beam spot is about 5 × 7 μm in size. We collected XRD patterns at several different positions including the center and edge of the sample chamber, to check the data consistency. A symmetric diamond anvil cell (DAC) with an opening angle of 70° and two pieces of boehler-type diamonds with culet of 300 μm in diameter was employed to generate high pressure. The distance between sample and detector was calibrated by the $CeO_2$ standard. The intensity vs two-theta patterns were integrated using the Dioptas software package.

**Computational details**

The electronic properties of Na-GIC were investigated using the Vienna *ab initio* simulation package within the framework of density functional theory (*50*), employing Perdew–Burke–Ernzerhof generalized gradient approximation (*51*). The interactions between electrons and ions were described using projector augmented wave (PAW) method (*52*), with $2s^22p^63s^1$ and $2s^22p^2$ valence electrons for Na and C atoms, respectively. A plane-wave basis set with an energy cutoff of 1000 eV and uniform $\Gamma$-centered $k$-point grids with a resolution of $2\pi \times 0.022$ Å$^{-1}$ were employed for electronic self-consistent calculations. Phonon dispersion and electron-phonon coupling (EPC) were calculated using density functional perturbation theory as implemented in the QUANTUM ESPRESSO package (*53*). Optimized norm-conserving Vanderbilt pseudopotentials were used (*54*), with a kinetic energy cutoff of 100 Ry and a charge density cutoff of 400 Ry. Given the layered structure of $NaC_8$, the van der Waals (vdW) interactions are included in the calculations by using the vdW-DF-obk8 functional (*55*). Self-consistent electron density and EPC were calculated employing 24 × 12 × 20 $k$-point meshes and 6 × 3 × 5 $q$-point meshes for the investigated structure. The superconducting transition temperature ($T_c$) was evaluated by using the Allen-Dynes-modified McMillan formula (*56, 57*),

$$T_c = \frac{\omega_{\log}}{1.2}\exp\left[-\frac{1.04(1+\lambda)}{\lambda-\mu^*(1+0.62\lambda)}\right],$$

where $\lambda$ is the EPC constant, $\omega_{\log}$ is the logarithmic average phonon frequency, and $\mu^*$ is the Coulomb pseudopotential parameter. The parameters $\lambda$ and $\omega_{\log}$ are defined as

$$\lambda = 2\int_0^\infty \frac{\alpha^2 F(\omega)}{\omega}d\omega,$$

and

$$\omega_{\log} = \exp\left[\frac{2}{\lambda}\int_0^\infty \frac{d\omega}{\omega}\alpha^2 F(\omega)\ln\omega\right].$$

Maximally localized Wannier functions (MLWFs) (*58*) were constructed using a 6 × 3 × 5 grid of the Brillouin zone. The anisotropic Migdal-Eliashberg equations (*59, 60*) were solved on fine uniform 48 × 24 × 40 $k$- and 24 × 12 × 20 $q$-point grids, with an energy window of -0.13~0.25 eV around the Fermi level and a Matsubara frequency cutoff of 1.0 eV. For solving the isotropic ME equations, the Dirac δ functions for electrons and phonons were broadened using Gaussians smearing with widths of 0.05 meV and 0.02 meV, respectively. The Matsubara frequency was truncated at $\omega_c$ = 1 eV, approximately sixfold compared to that of the highest phonon excitation energy.

**Acknowledgments:** We thank the Shanghai Synchrotron Radiation Facility of BL17UM (31124. 02. SSRF.BL17UM) for the assistance on synchrotron XRD measurements. **Funding:** This work was supported by the National Key R&D Program of China (Grant Nos. 2022YFA1402300 and 2023YFA1406200), National Natural Science Foundation of China (Grant Nos. 52025026, 52288102, 52090020, 52372157), and the S&T Program of Hebei (225A1102D). **Author contributions:** M.X.H., Y.Q.L. and C.M.H. contributed equally to this work. X.F.Z. conceived the project. M.X.H., Y.Q.L., S.W. and F.K. prepared the samples and conducted the experiments under L.W. and X.F.Z's supervision. C.M.H., F.K. and X.F.Z. analyzed the crystal structure. C.M.H., T.A., X.S., and G.Y. performed the theoretical calculations under X.F.Z's supervision. F.K., B.X., X.F.Z. and Y.T. drafted the manuscript with input from all authors. All authors contributed to the discussion and revision of the paper.

**Competing interests:** Authors declare that they have no competing interests.

**Data and materials availability:** All data needed to evaluate the conclusions in the paper are present in the paper and/or the Supplementary Materials.

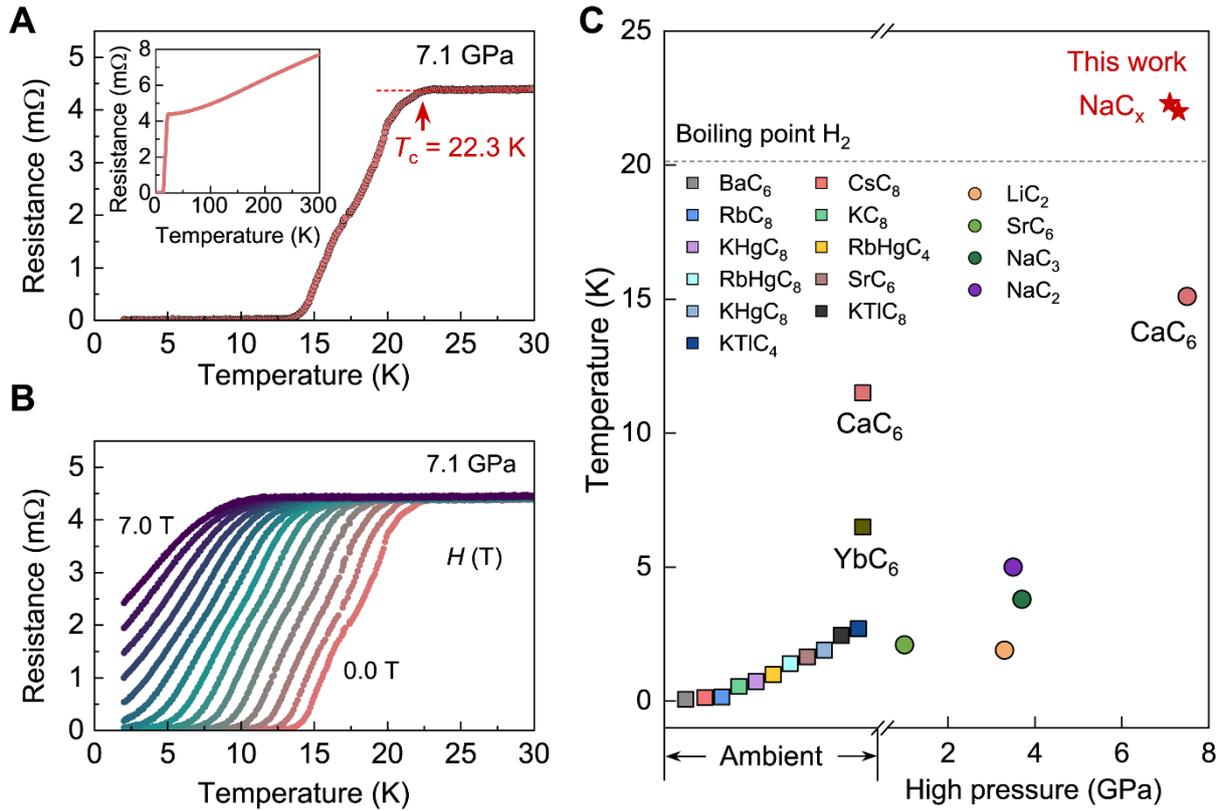

**Fig. 1. Superconductivity in Na-intercalated graphite under compression.** (A), Temperature-dependent electrical resistance of Na-intercalated graphite at 7.1 GPa, showing superconducting transition ($T_c$) at approximately 22.3 K. (B), Suppression of $T_c$ in Na-intercalated graphite under increasing external magnetic fields. (C), Comparison of $T_c$ between our Na-intercalated graphite and previously reported superconducting GICs at ambient- or high-pressure conditions.

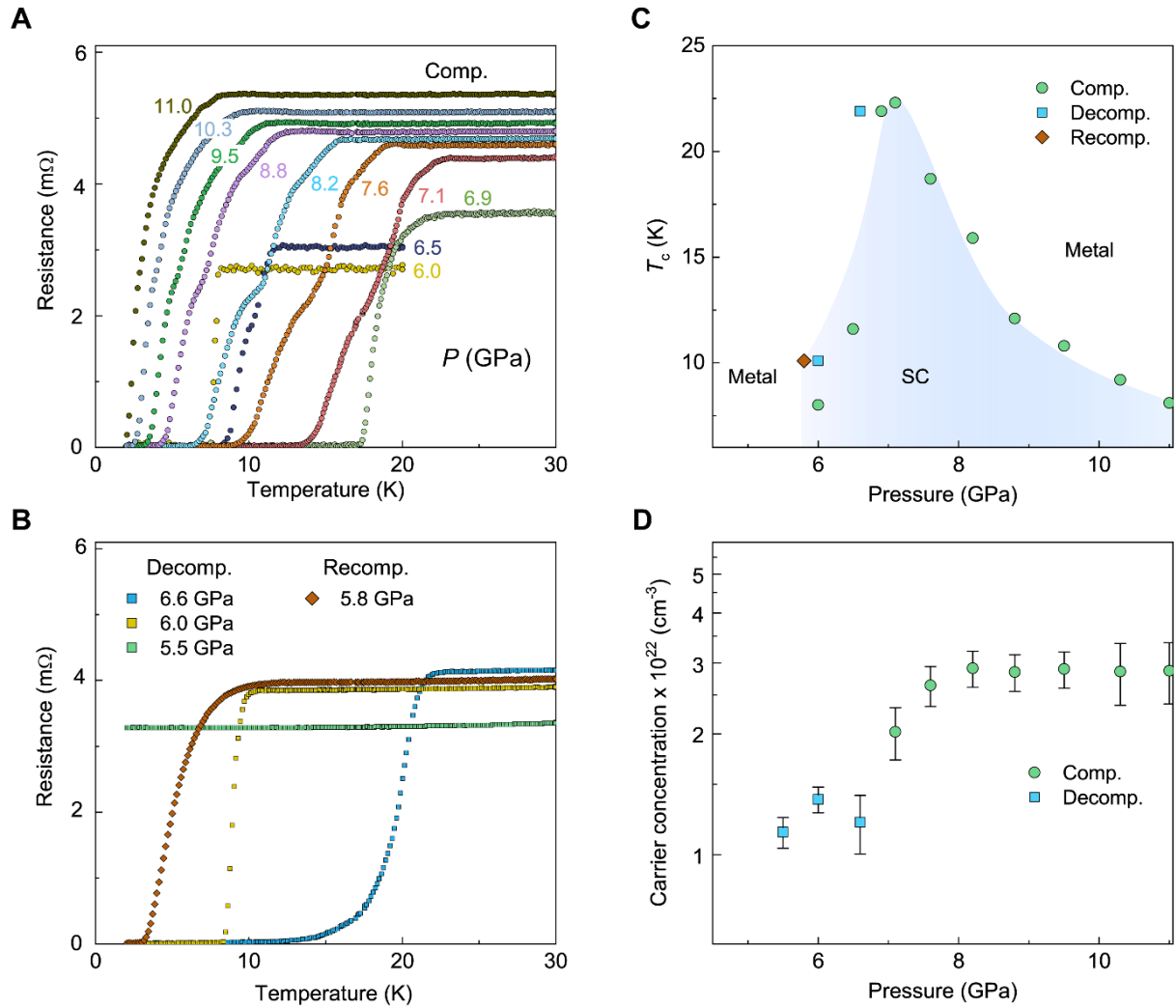

**Fig. 2. Comprehensive electrical transport results of Na-intercalated graphite through pressure cycling.** (A), Temperature-dependent resistance curves obtained during compression (Comp.), showing the pressure-dependent superconductivity. (B), Resistance–temperature curves during decompression (Decomp.) and subsequent recompression (Recomp.) cycle, demonstrating the reversible feature of the pressure-induced superconducting order. (C), Superconducting phase diagram mapping $T_c$ versus $P$ up to 11 GPa. (D), Evolution of carrier concentrations as a function of applied pressure at 30 K, with error bars derived from Hall coefficient measurements.

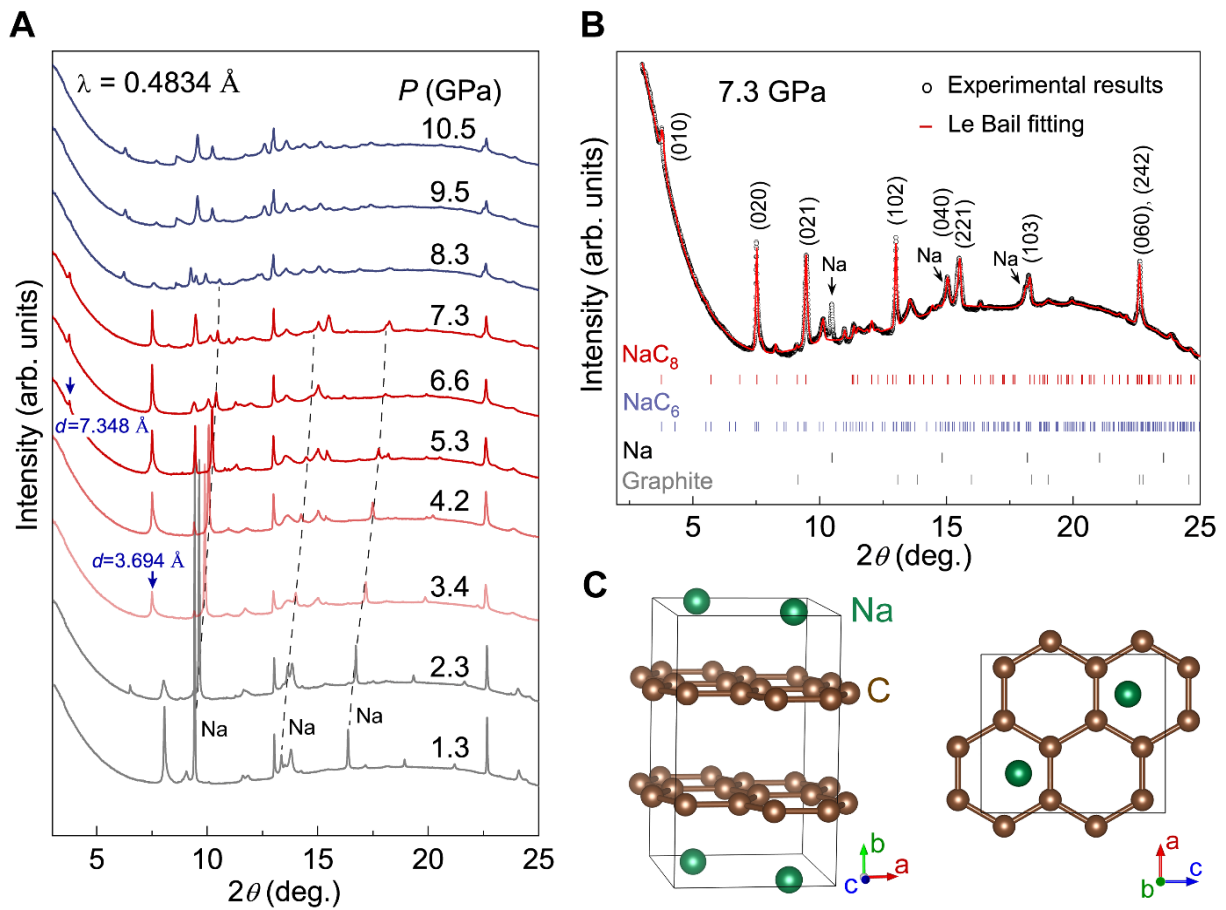

**Fig. 3. Crystal structure characterization of Na-intercalated graphite under compression.** (A), Room-temperature XRD patterns collected with increasing pressure. The blue arrows highlight the emergence of new diffraction peaks at elevated pressures, indicative of the formation of a few intercalations. (B), Le bail fitting of the XRD patterns at 7.3 GPa (open circles: experimental data; red line: fitting pattern) using the orthorhombic stage-2 $NaC_8$ with *Pmma* and $NaC_6$ with *P2/m* symmetry. The red, blue, black, and gray vertical ticks indicate the indexed peaks of $NaC_8$, $NaC_6$, Na metal, and graphite, respectively. (C), Crystal structure of $NaC_8$ showing the characteristic stage-2 intercalation arrangement. Na and C atoms are colored in green and brown, respectively.

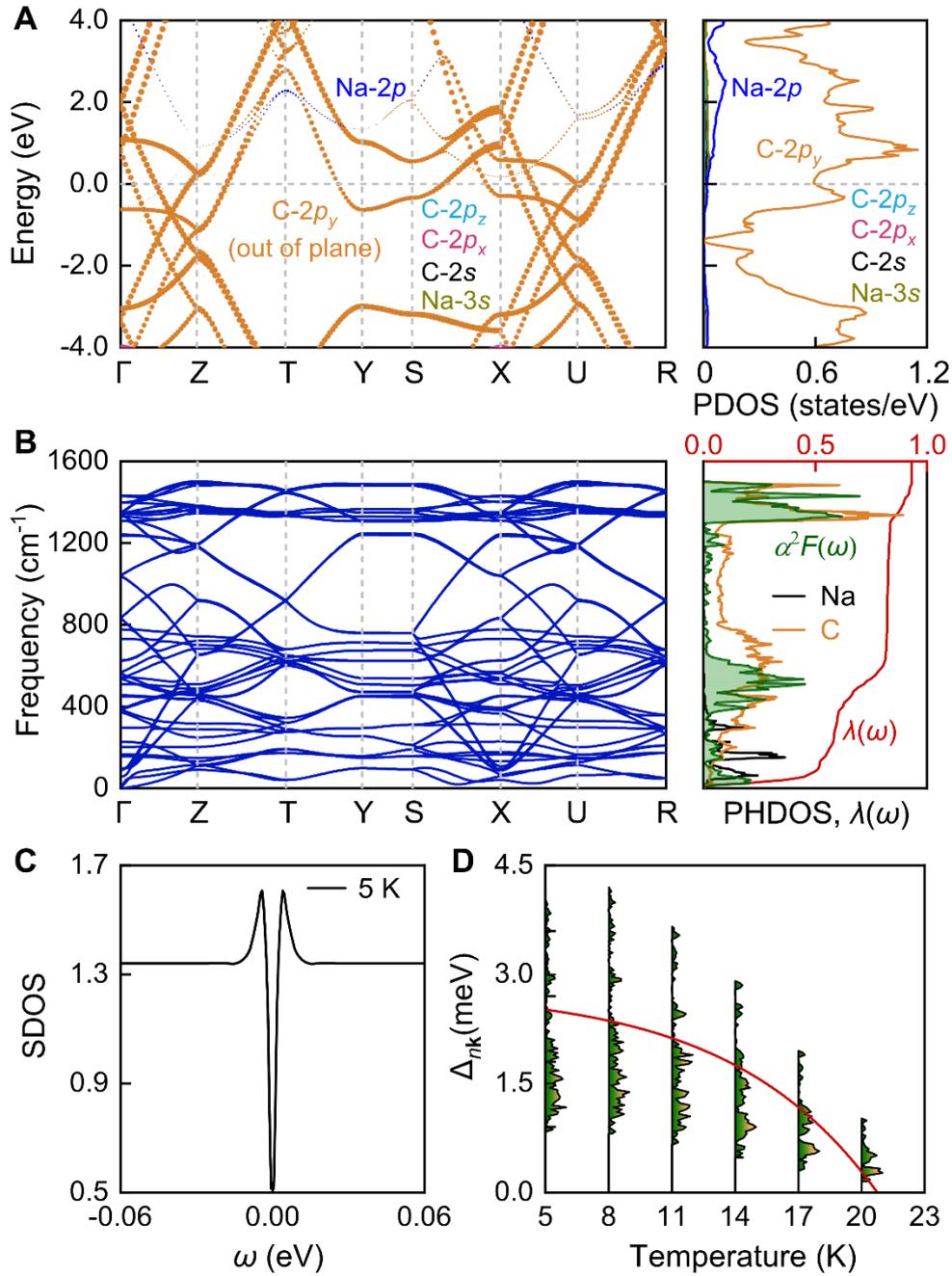

**Fig. 4. Electronic structure and EPC analysis of doped NaC$_8$ at 7 GPa with an electron doping concentration of 0.02 $e$/atom.** (A), Orbital-resolved electronic band structure and projected density of states (PDOS). Notice that the lattice vector ***b*** is oriented along the stacking direction. (B), Phonon dispersion alongside the corresponding phonon density of states (PHDOS), Eliashberg spectral function $\alpha^2 F(\omega)$, and the accumulated EPC $\lambda(\omega)$. (C), Renormalized superconducting density of states (SDOS) calculated at 5 K. (D), Temperature evolution of the anisotropic superconducting gap $\Delta_{n\mathbf{k}}$, predicting a $T_c$ of 20.8 K.

# Supplementary Materials for

# **Superconductivity at 22.3 K in Compressed Sodium-intercalated Graphite**

Ming-Xing Huang *et al.*

*Corresponding author. Email: fengke@ysu.edu.cn, xfzhou@ysu.edu.cn, fhcl@ysu.edu.cn

**This file includes:**

    Discussion Section I and Section II
    figs. S1 to S11
    Tables S1
    References (1 to 5)

**Discussion**

**Section I: Determination of stage index (*n*)**

We conducted quantitative stage analysis using characteristic (00*l*) reflections, following established methodologies from published studies (*1-3*),

$$d_{00l} = I_c / l \quad \text{and} \quad d_{00l+1} = I_c / (l+1) \quad (1)$$

$$I_c = d_s + (n-1)d_g \quad (2)$$

where $d_{00l}$ and $d_{00l+1}$ are the d-spacings of (00*l*) and (00*l*+1) reflections, and $I_c$, $d_s$ and $d_g$ is the *c*-axis repeating distance, intercalation sandwich thickness and graphite interlayer spacing, respectively. For the ambient-pressure sample, adjacent (00*l*) and (00*l*+1) peaks appear at ~25.4° ($d \approx 3.504$ Å) and ~28.6° ($d \approx 3.119$ Å), as shown in fig. S1. Based on the Eqn. 1, $d_{00l} \times l = d_{00l+1} \times (l+1)$, we obtain $l/(l+1) = d_{00l+1}/d_{00l} \approx 8/9$, yielding $l \approx 8$. The *c*-axis repeating distance $I_c = d_{008} \times 8 \approx 28.032$ Å. Given that the Na intercalation sandwich thickness $d_s \approx 4.50$ Å (*4*) and graphite interlayer spacing $d_g \approx 3.35$ Å, the stage index (*n*) is evaluated as $n = (I_c - d_s)/d_g + 1 \approx 8$, confirming a stage-8 Na-GIC consistent previously published study (*5*).

Similarly, using $d_s \approx 4.50$ Å, $d_g \approx 3.35$ Å and $I_c = d_s + (n-1)d_g$, we evaluated $I_c$ is approximately 4.50, 7.85 and 11.20 Å for stage-1, stage-2 and stage-3 Na-GICs, respectively. At 7.3 GPa, characteristic peaks appear at ~3.77° ($d \approx 7.348$ Å) and ~7.50° ($d \approx 3.695$ Å), as shown in Fig. 3a. Similar analysis yields $l/(l+1) \approx 1/2$, giving $l \approx 1$ and $I_c \approx 7.348$ Å. This value closely matches that of the stage-2 Na-GIC calculated above ($I_c \approx 7.85$ Å) when accounting for compression effects, confirming the formation of a stage-2 Na-GIC at 7.3 GPa.

**Section II: Structure identification of the major *Pmma* and minor *P2/m* phases**

The quantitative intercalation stage analysis (section I) demonstrates that the superconducting phase originates from stage-2 Na-GICs. We then constructed various stage-2 candidates and refined with the experimental XRD patterns. Our refinement reveals that a stage-2 $NaC_8$ structure with *Pmma* symmetry provides the best match to our experimental data (fig. S6), while several weak diffraction peaks at ~10.1°, ~10.5° and ~11.0° deviate from the simulated XRD pattern. The ~10.5° peak arises from the residual Na. Through systematic peak analysis, we identified the unindexed reflections at ~10.1° and ~11.0° is consistent with a stage-2 $NaC_6$ structure with *P2/m* symmetry. The simulated pattern of stage-2 $NaC_6$ matches both the peaks at ~10.1° and ~11.0° and the characteristic stage-2 Na-GIC peaks at ~3.77° and ~7.50° (fig. S6A). Incorporating this two-phase model yields substantially better agreement with the experimental data (fig. S6B), demonstrating that the measured samples consist of coexisting major *Pmma* and minor *P2/m* phases. The fitted lattice parameters of the minor *P2/m* phase at 7.3 GPa are $a$ = 5.079±0.011 Å, $b$ = 7.364±0.018 Å, $c$ = 6.544±0.012 Å, and $\beta$= 98.8°±2.5°. Carbon atoms occupy the positions (from calculations) of (0.278, 0.194, 0.611), (0.389, 0.196, 0.056), (0.556, 0.192, 0.722), (0.222, 0.185, 0.389), (0.111, 0.194, 0.944) and (0.945, 0.192, 0.278), while sodium atoms occupy the (0.759, 0.500, 0.645) and (0.280, 0.500, 0.847) positions, respectively. Some XRD peaks of $NaC_8$ (or $NaC_6$) show higher or lower intensities compared to the simulated patterns, which may arise from preferred orientation or texturing, as evidenced by the distinctive spot-like features (fig. S8), which fundamentally alters peak intensities.

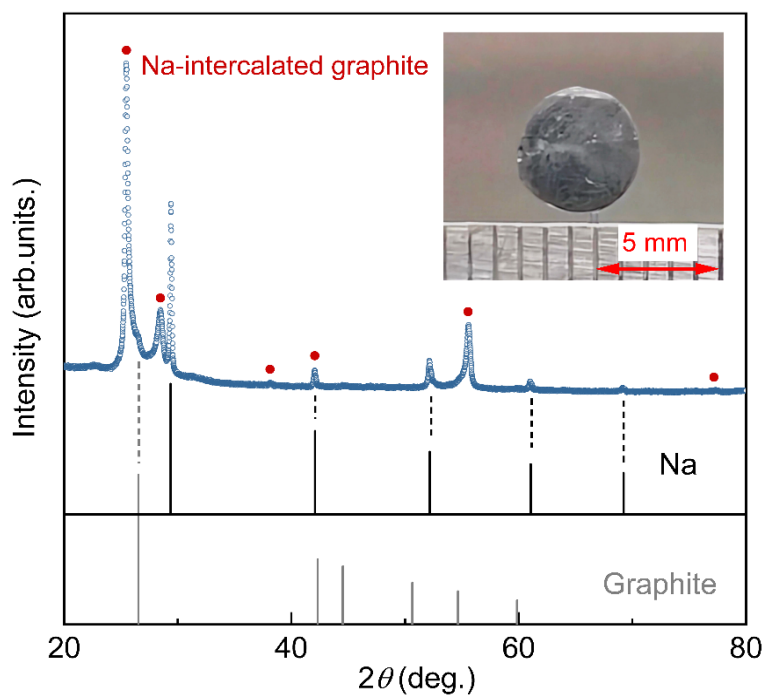

**Fig. S1.** Ambient-pressure XRD pattern of the as-synthesized Na-GIC after grinding the mixture of sodium metal and graphite by half an hour. The inset shows an optical photograph of the as-formed samples. The XRD result shows that a certain amount of sodium (Na) and graphite raw materials remain unreacted after grinding. These residual Na can act as an internal source material for further interaction under compression. The ambient-pressure XRD pattern reveals that the synthesized Na-GICs is stage-8 $NaC_{64}$ (section I), consistent with previously published results [5].

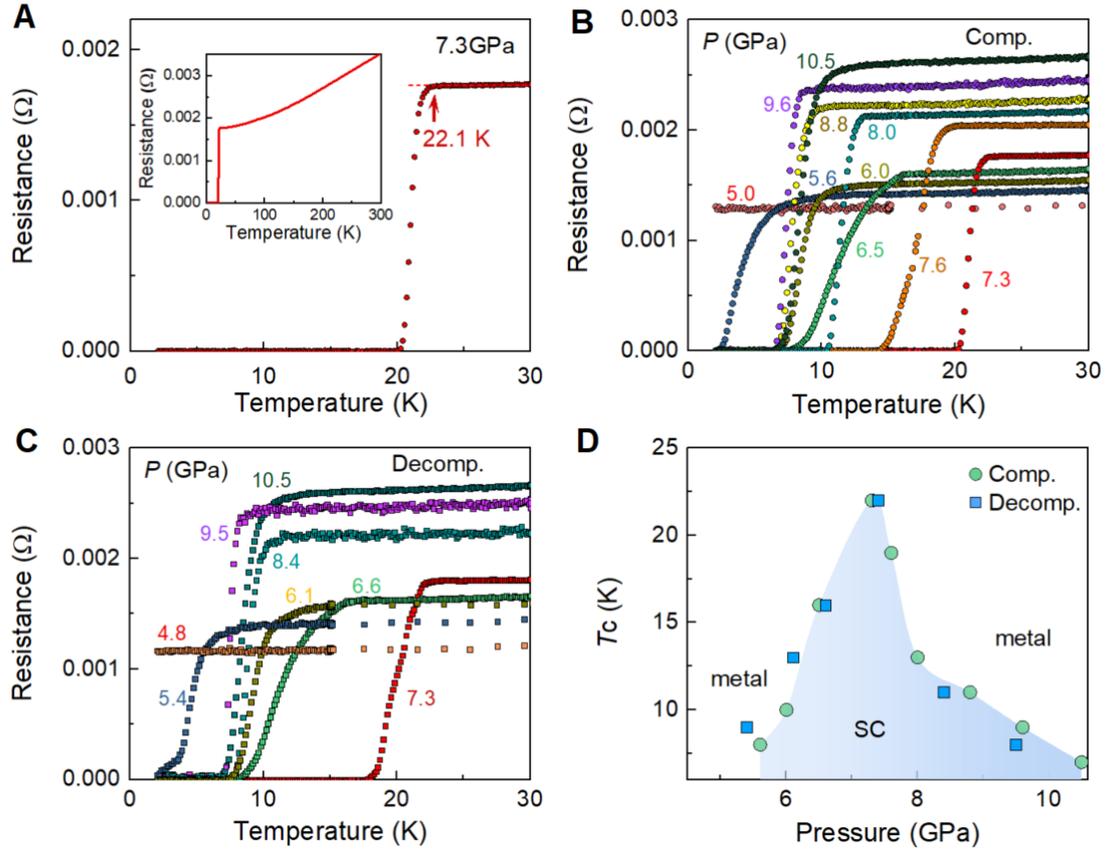

**Fig. S2. Superconductivity in an additional Na-GIC sample under compression.** (A), Temperature-dependent electrical resistance at 7.3 GPa, showing superconducting transition ($T_c$) at ~22.1 K. (B) and (C), Resistance – temperature curves at representative pressures during compression (Comp.) and decompression (Decomp.) cycle. (D), Superconducting $T_c$ – $P$ phase diagram showing the reversible behavior of superconducting transition. We notice that fig. S2 reveals a broadened transition at 6.5 GPa but sharpens distinctly at 7.3 GPa. In contrast, the superconducting transition is sharp at 6.9 GPa but broadens significantly at 7.1 GPa in Fig. 2A. Moreover, these broadening effects prove irreversible upon decompression. These differences may arise from local sample inhomogeneity caused by local pressure gradient and mechanical stress within the diamond anvil cell, rather than intrinsic Na-GIC behavior. Pressure gradient of ~0.5 GPa exist across the sample chamber.

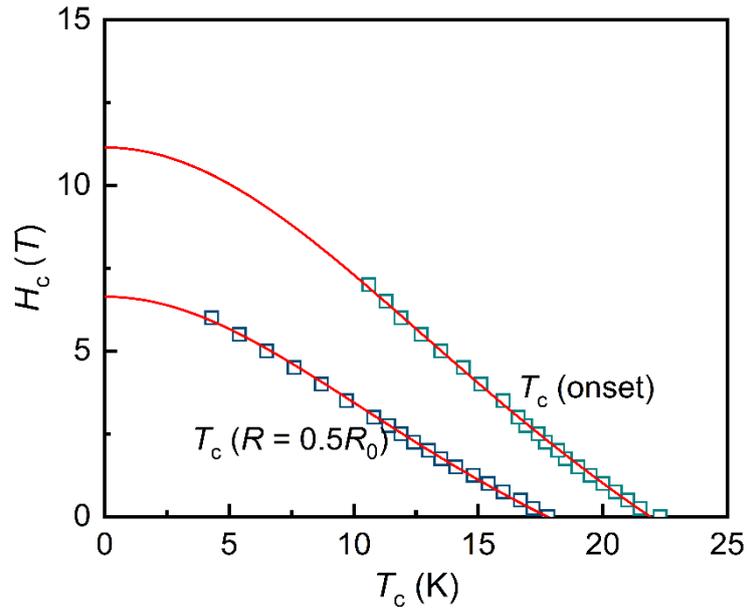

**Fig. S3**. Ginzburg–Landau fitting of the $H_c(T)$ vs $T_c$ curves. $T_c$ was defined as both the onset temperature of superconducting transition [$T_c$(onset)] and half of the normal-state resistances before accessing the superconducting states [$T_c(R = 0.5R_0)$].

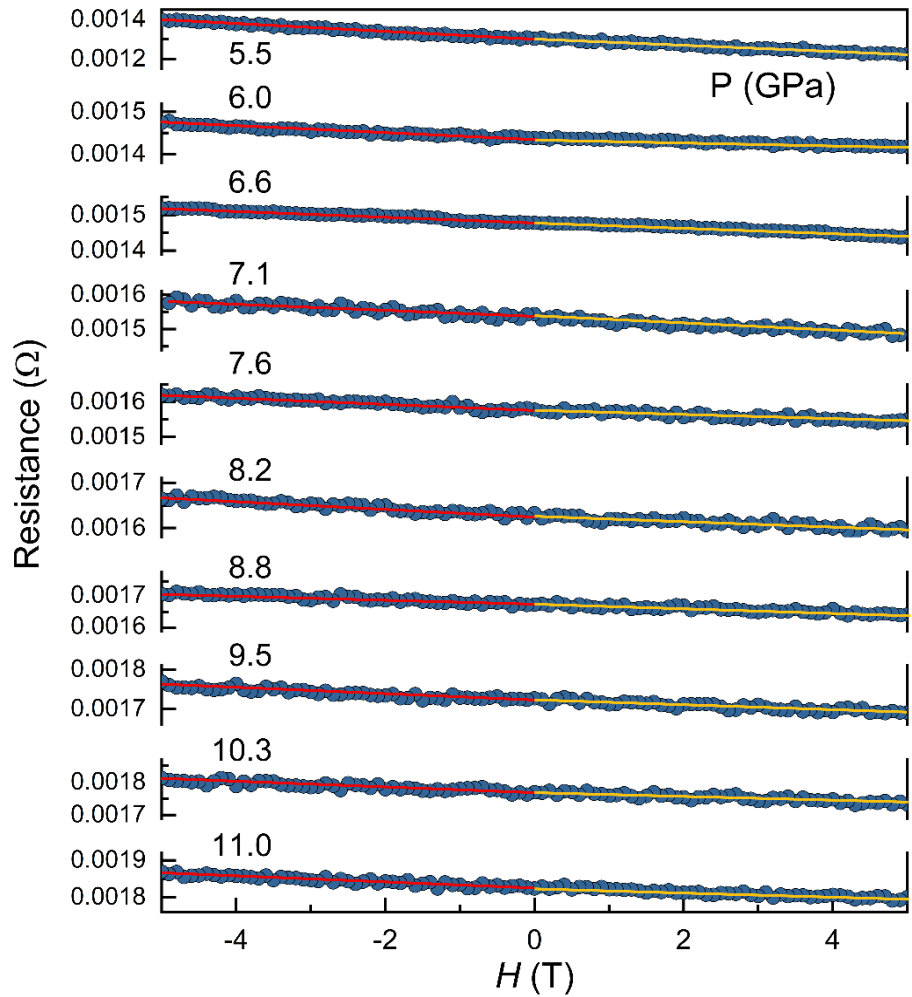

**Fig. S4**. Raw data of Hall effect measurements on Na-GIC under compression. Magnetic fields (-5 – 5 T) were applied perpendicular to the surface of diamond anvil cell. The red and yellow lines are the linear fits of the data from -5 to 0 T and 0 to 5 T, respectively. Hall resistances were then obtained by subtracting the interpolated resistance values at negative magnetic fields from those at related positive magnetic fields to exclude the asymmetric effect. By analyzing the Hall data, negative Hall coefficients were obtained, indicating that Na-GIC hosts electron-type carriers under compression.

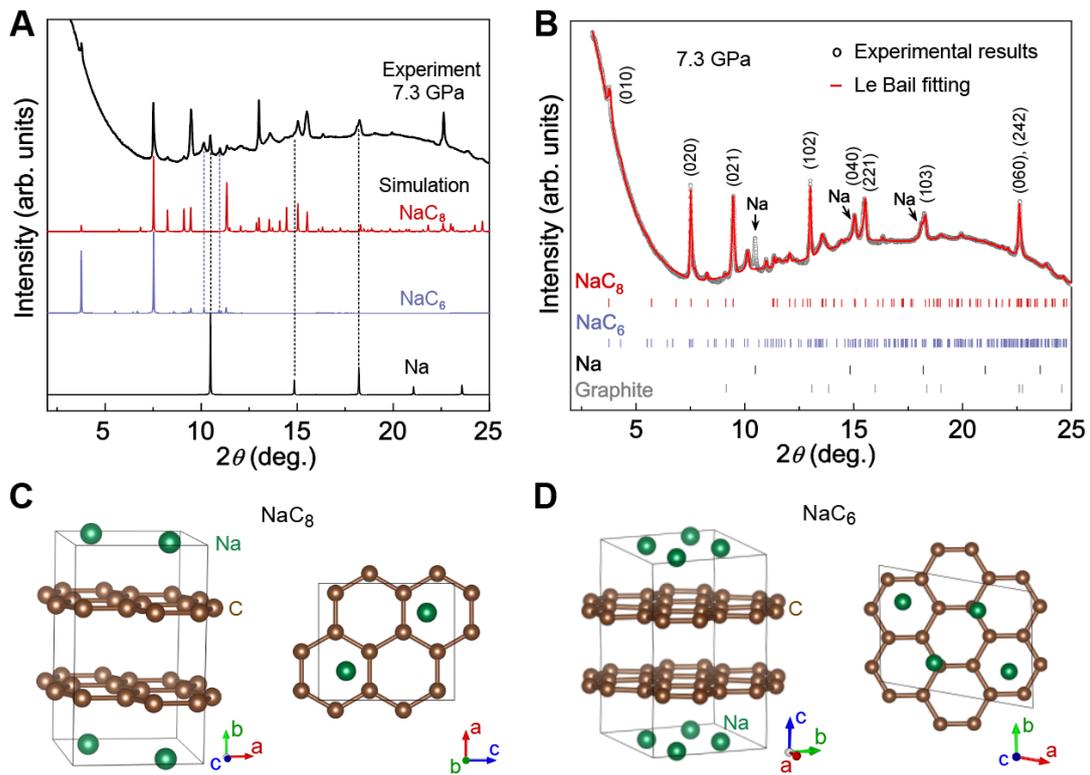

**Fig. S5. Structural characterization and phase identification of high-pressure Na-GIC.** (A) Simulated XRD patterns for the major *Pmma* NaC$_8$ and minor *P2/m* NaC$_6$ phases. (B) Le Bail refinement results using both the *Pmma* and *P2/m* structures. (C) and (D), Crystalline structures of the major *Pmma* and minor *P2/m* phases, respectively.

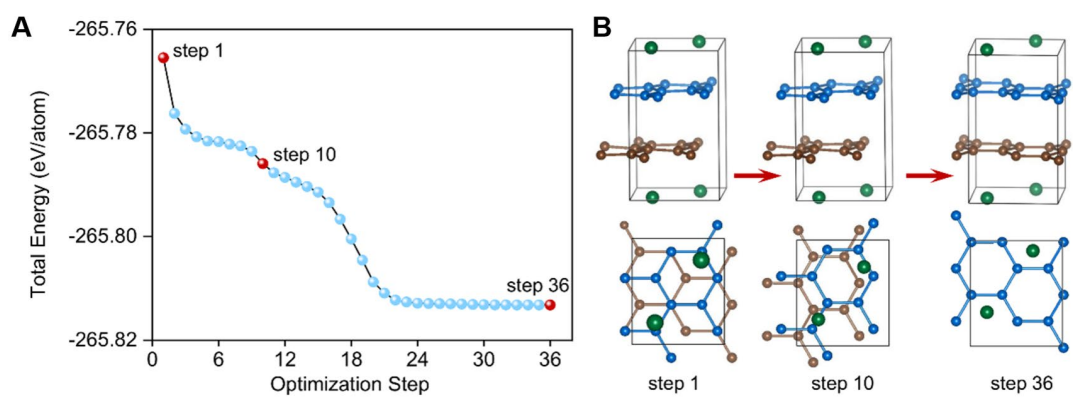

**Fig. S6. Theoretical validation of AA-stacking preference in high-pressure NaC$_8$.** Energy convergence (A) and structural evolution (B) of AB-stacked NaC$_8$ at 7 GPa. The results demonstrate that AA-stacking NaC$_8$ achieves lower total energy and better structural stability than the AB-stacking counterpart, confirming the thermodynamic preference for AA-stacking under compression. The transition from conventional AB to AA stacking can be rationalized as follows: under compression, sodium intercalation significantly alters the electronic structure through substantial charge transfer from sodium atoms to graphene sheets. This charge redistribution modifies the electrostatic environment and van der Waals interactions between layers, ultimately favoring AA-stacking over conventional arrangements.

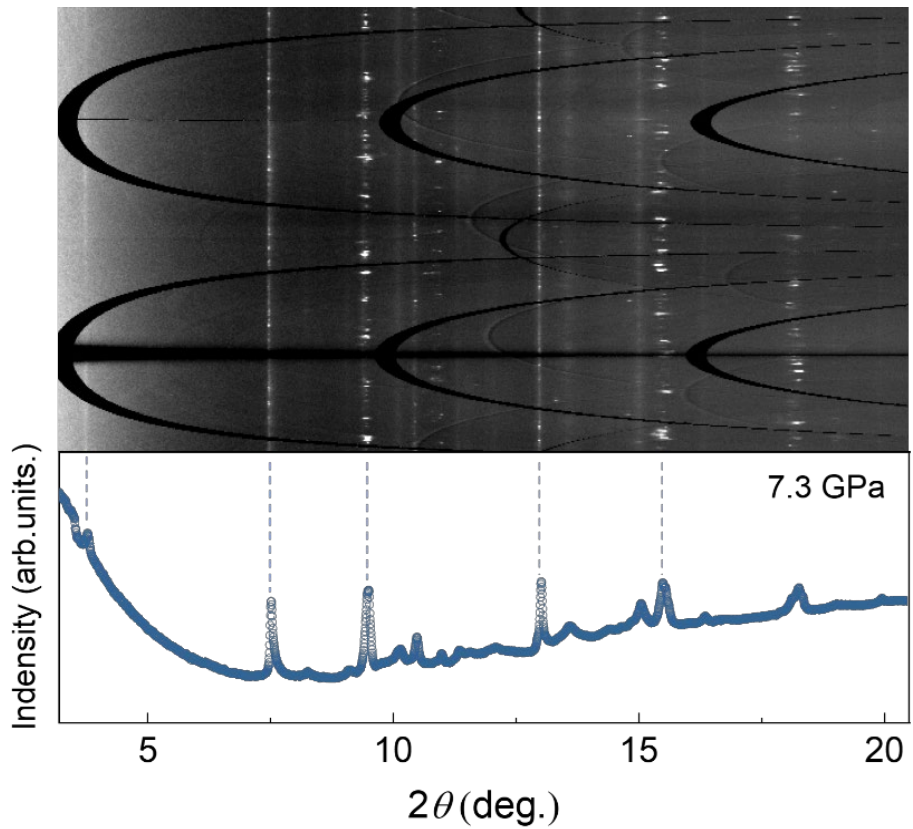

**Fig. S7**. Raw data of the synchrotron X-ray diffraction (XRD) experiments on Na-GIC at 7.3 GPa. The XRD pattern is caked for clear comparison. The XRD pattern shows spot-like features, which are likely caused by preferred orientations (texting) within the sample under compression.

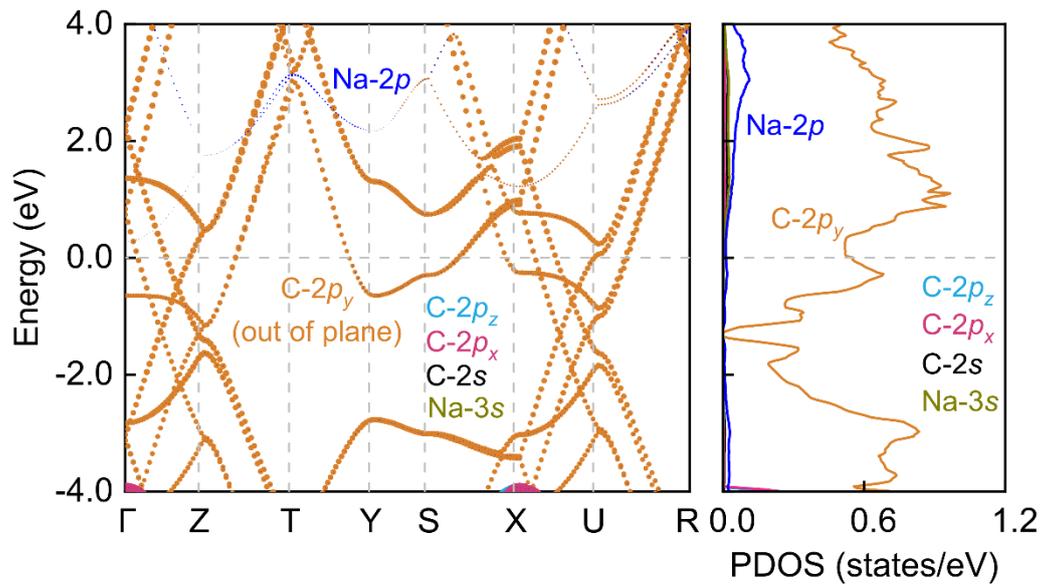

**Fig. S8.** Orbital-resolved band structures and projected density of states (PDOS) of intrinsic $NaC_8$ at 7 GPa.

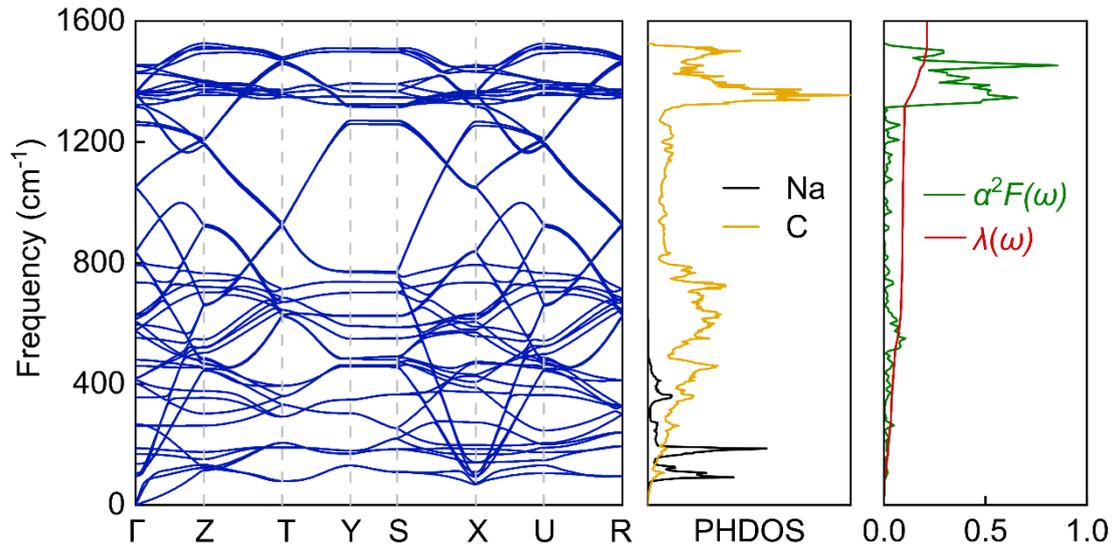

**Fig. S9.** Phonon dispersion, phonon density of states (PHDOS), Eliashberg spectral function $\alpha^2F(\omega)$, and accumulated EPC $\lambda(\omega)$ for intrinsic $NaC_8$ at 7 GPa.

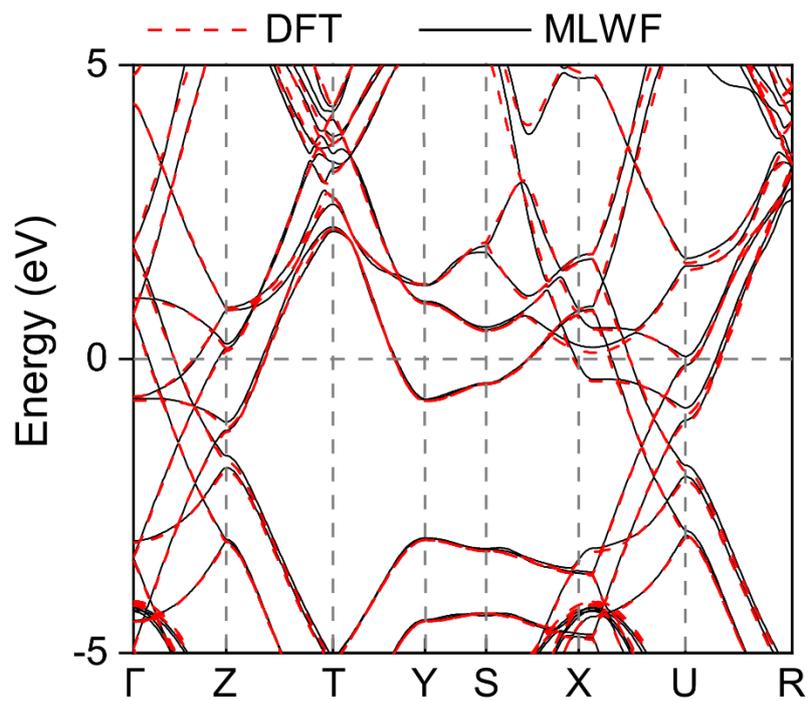

**Fig. S10.** Band structure of NaC$_8$ with the electron doping concentration of 0.02 *e*/atom at 7 GPa. The red and black lines represent the band structures obtained by the density functional theory calculation and the MLWF interpolation, respectively.

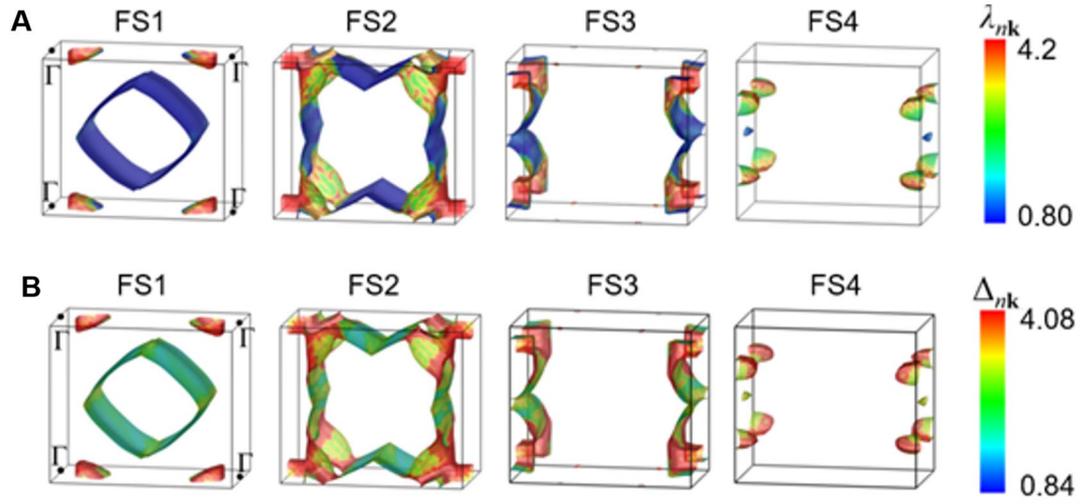

**Fig. S11.** Momentum-resolved (A) EPC strength $\lambda_{n\mathbf{k}}$ and (B) superconducting gap $\Delta_{n\mathbf{k}}$ projected on four FS sheets of NaC$_8$ at 5 K, with the electron doping concentration of 0.02 $e$/atom at 7 GPa.

**Table S1**. The calculated lattice constants, logarithmic averaged phonon frequency ($\omega_{\log}$), EPC constant ($\lambda$), and the estimated superconducting critical temperature ($T_c$) using the Allen-Dynes-modified McMillan equation with a $\mu^* = 0.1$ for $NaC_8$ under various doping concentrations and pressures. Note that vdW interaction is included for all calculations.

| Pressure (GPa) | Doping (e/atom) | Lattice parameters (Å) | | | $\omega_{\log}$ (K) | $\lambda$ | $T_c$ (K) |
|---|---|---|---|---|---|---|---|
| | | a | b | c | | | |
| 6 | 0 | 4.290 | 7.237 | 4.950 | 1012.01 | 0.25 | 0.05 |
| 7 | 0 | 4.286 | 7.190 | 4.946 | 1073.96 | 0.21 | 0.002 |
| 7 | 0.01 | 4.295 | 7.287 | 4.956 | 749.35 | 0.35 | 1.40 |
| 7 | 0.02 | 4.296 | 7.538 | 4.958 | 175.47 | 0.93 | 10.88 |
| 7 | 0.03 | 4.271 | 8.602 | 4.935 | - | - | - |